\documentclass[10pt, journal, lettersize]{IEEEtran}
\IEEEoverridecommandlockouts
\usepackage{xspace,amsmath,amssymb,amsfonts,epsfig,subfigure,syntonly}
\usepackage[skip=3pt]{caption}
\usepackage{lipsum}
\usepackage{cite,bm,color,url,textcomp,empheq,boxedminipage}
\usepackage{algorithmicx}
\usepackage{indentfirst}
\usepackage{epstopdf,makecell}
\usepackage{empheq}
\usepackage{multicol}
\usepackage{pifont}
\usepackage{cuted}  
\usepackage{stfloats}
\usepackage[noend]{algpseudocode}
\usepackage{tcolorbox}
\usepackage[linesnumbered,ruled]{algorithm2e}
\usepackage{graphicx,graphics}  
\usepackage{multirow,multicol}
\usepackage{psfrag}    
\usepackage{stfloats}
\usepackage{url}
\usepackage{lipsum}

\newcommand\blfootnote[1]{%
  \begingroup
  \renewcommand\thefootnote{}\footnote{#1}%
  \addtocounter{footnote}{-1}%
  \endgroup
}

\long\def\symbolfootnote[#1]#2{\begingroup
\def\thefootnote{\fnsymbol{footnote}}
\footnote[#1]{#2}\endgroup}
\psfull

\allowdisplaybreaks[4]
\title{\Huge MIMO Pinching-Antenna-Aided SWIPT}
\author{Haoyun Li, Zhonghao Lyu, \emph{Member, IEEE}, Yulan Gao, \emph{Member, IEEE}, \\ Ming Xiao, \emph{Senior Member, IEEE}, and H. Vincent Poor, \emph{Life Fellow, IEEE}}

\begin{document}
\maketitle
\begin{abstract}
    Pinching-antenna systems (PASS) have recently emerged as a promising technology for improving wireless communications by establishing or strengthening reliable line-of-sight (LoS) links by adjusting the positions of pinching antennas (PAs). Motivated by these benefits, we propose a novel PASS-aided multi-input multi-output (MIMO) system for simultaneous wireless information and power transfer (SWIPT), where the PASS are equipped with multiple waveguides to provide information transmission and wireless power transfer (WPT) for several multiple antenna information decoding receivers (IDRs), and energy harvesting receivers (EHRs), respectively. Based on the system, we consider maximizing the sum-rate of all IDRs while guaranteeing the minimum harvested energy of each EHR by jointly optimizing the pinching beamforming and the PA positions. To solve this highly non-convex problem, we iteratively optimize the pinching beamforming based on a weighted minimum mean-squared-error (WMMSE) method and update the PA positions with a Gauss-Seidel-based approach in an alternating optimization (AO) framework. Numerical results verify the significant superiority of the PASS compared with conventional designs.
\end{abstract}
\blfootnote{H. Li, Z. Lyu, Y. Gao, and M. Xiao are with the Division of Information Science and Engineering, KTH Royal Institute of Technology, Stockholm 10044, Sweden (email: \{haoyunl, lzhon, yulang, mingx\}@kth.se).

H. V. Poor is with the Department of Electrical and Computer Engineering, Princeton University, New Jersey 08544, USA (e-mail:
poor@princeton.edu).}

\section{Introduction}
The sixth-generation (6G) wireless networks are expected to achieve a higher transmission frequency and larger data rate with extremely low latency and hyper-reality compared with the fifth-generation (5G) technologies \cite{6G, you}. To support these novel features, a massive number of Internet-of-Things (IoT) devices, such as smartphones, wireless sensors, and wearable devices, have emerged to support 6G wireless networks, which require ubiquitous communication connectivity and a stable energy supply. To this end, simultaneous wireless information and power transfer (SWIPT) has recently been viewed as a viable solution by superimposing information and energy within the power domain of a unified signal \cite{6G3}. However, an energy harvesting receiver (EHR) typically requires much more receive power than that for an information decoding receiver (IDR) due to drastically decaying wireless power transfer (WPT) efficiency caused by long-distance propagation path loss. 

To tackle this challenge, previous works have explored various transmission architectures to improve SWIPT performance by increasing spatial degrees of freedom (DoFs), including multiple-input multiple-output (MIMO) \cite{MIMO}, intelligent reflecting surfaces (IRSs) \cite{IRS}, and fluid antenna (FA) systems \cite{FA1}. Specifically, \cite{MIMO} optimized transmit covariance matrices and time-splitting ratios for enhanced MIMO energy efficiency. IRSs in \cite{IRS} improved the WPT efficiency by adjusting the reflective element phase shifts. Furthermore, \cite{FA1} demonstrated that reconfigurable FA positions can enhance both energy harvesting and communication performance by adapting to the radio propagation environment.
Despite their advantages, these architectures have some limitations. Traditional MIMO and IRSs struggle to adapt to dynamic environments because of fixed antenna placements. FA systems offer limited antenna movement relative to the carrier frequency and face challenges with high implementation costs. To overcome these drawbacks, pinching-antenna systems (PASS), a dielectric waveguide-based leaky-wave antenna technology \cite{PASS0, PASS01}, have emerged as a promising reconfigurable solution. Different from MIMO, IRS, and FA systems, PASS are expected to provide wider coverage by flexibly changing the positions of multiple pinching antennas (PAs) on the waveguides, which have strong flexibility in establishing or strengthening line-of-sight (LoS) links \cite{PASS1}. By optimizing the positions and numbers of PAs, PASS can enable a highly flexible and scalable pinching beamforming gain. Moreover, PASS can cover users within the near-field region due to the extended aperture of waveguides. Existing research has verified the superiority of PASS in multi-user MIMO \cite{PASS1}, integrated sensing and communications (ISAC) \cite{PASS2}, over-the-air computation \cite{PASS3}, and non-orthogonal multiple access (NOMA) \cite{PASS4}, etc. 
\begin{figure}[t]
    \centering
    \includegraphics[width=0.9\linewidth]{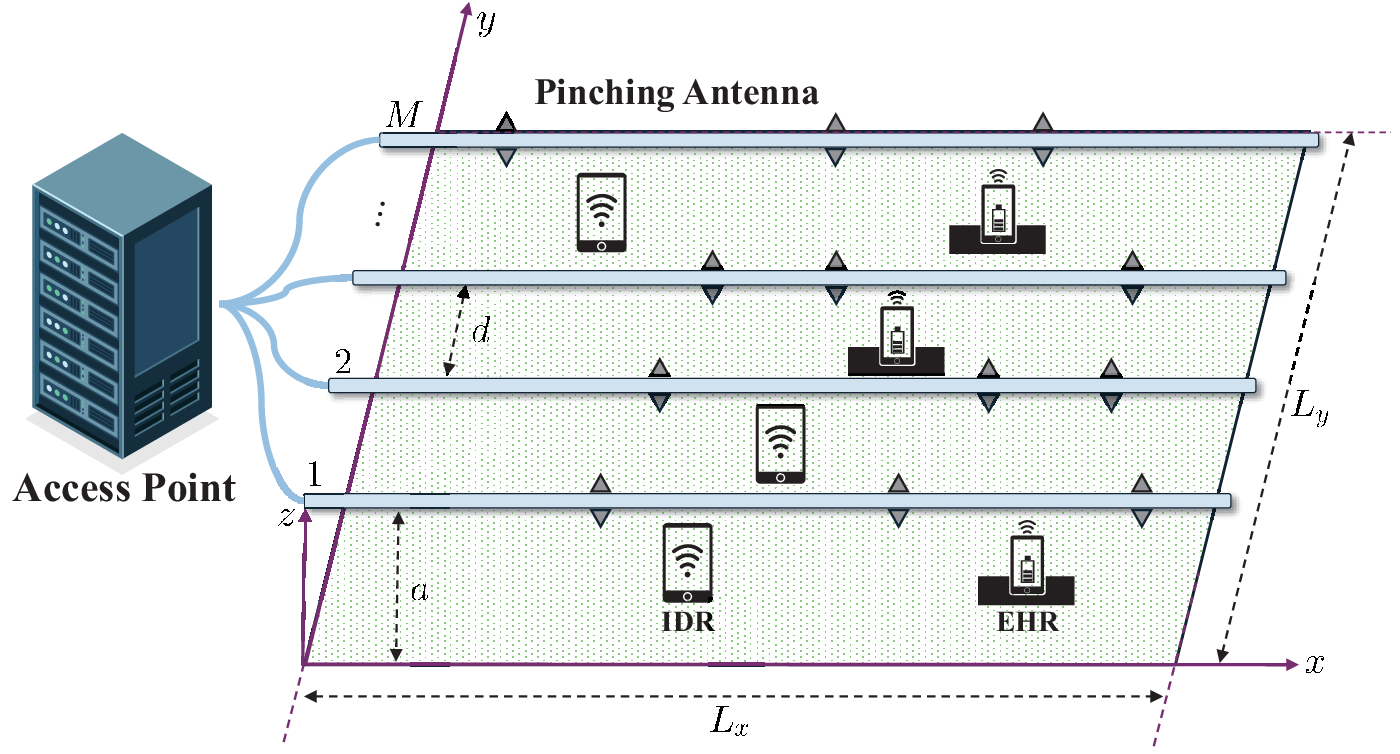}
    \caption{The illustration of a PASS-aided multi-user SWIPT system.}
    \label{system}
    \vspace{-10pt}
\end{figure}

Motivated by the significant potential of PASS, we investigate a novel downlink MIMO PASS-aided SWIPT system, which, to the best of our knowledge, has not been explored in the literature before. In the system, an access point (AP) applies a PASS consisting of multiple dielectric waveguides with multiple PAs activated on each waveguide to simultaneously transmit information and wireless power for multiple IDRs and EHRs, respectively. Each IDR and EHR is assumed to be equipped with multiple antennas. We aim to maximize the sum-rate of all IDRs while satisfying the minimum energy harvesting requirements of each EHR by jointly optimizing the pinching beamforming and the PA positions. We propose an alternating optimization (AO) framework to efficiently optimize the highly non-convex problem by combining a weighted minimum mean-squared-error (WMMSE)-based beamforming optimization method, together with a Gauss-Seidel-based PA position adjustment approach. Numerical results verify the superiority of the proposed design compared to the benchmark schemes.

\section{System Model and Problem Formulation}

In this work, we consider a MIMO PASS-aided SWIPT system where the PASS are fed with an AP to serve $K$ IDRs and $Q$ EHRs as illustrated in Fig. \ref{system}. The MIMO PASS system consists of $M$ parallel dielectric waveguides, each equipped with $N$ PAs. Considering a three-dimensional (3D) coordinate system, we assume that all waveguides of the PASS extend over the $x$-axis and are parallel over the $y$-axis at the altitude of $a$. The spacing between each two waveguides is $d$.  The position of the $n$-th PA on the $m$-th waveguide is given by $\mathbf{v}_{m,n}=[l_{m,n}, (m-1)d,a]$, where $0\leq l_{m,n}\leq L_x$ with $L_x$ denoting the length of each waveguide. For each waveguide $m$, the signal is fed at $[0, (m-1)d, a]$. We assume that all IDRs and EHRs are distributed over a rectangular area on the $x$-$y$ plane with sides $L_x$ and $L_y$. Each IDR and EHR is assumed to be equipped with a uniform linear array (ULA) with $J$ antennas. The spacing between adjacent antenna elements of the ULA is denoted by $d_s$. The start point of the ULA of IDR $k$ is located at $\mathbf{u}_{k,0}=[x_k, y_k, 0]$, while that of EHR $q$ is located at $\mathbf{u}_{q,0}=[x_q, y_q, 0]$, respectively. We assume that all ULAs are extended over the $x$-axis. Then, the $j$-th antenna of IDR $k$ is located at $\mathbf{u}_{k,j}=[x_k+jd_s, y_k,0]$, while that of EHR $k$ is located at $\mathbf{u}_{q,j}=[x_q+jd_s, y_q,0]$, respectively. Without loss of generality, we assume the signal vector fed to the AP is defined as
\begin{equation}
    \mathbf{s}=\sum^{K}_{k=1}\mathbf{W}_k\mathbf{s}_k \in \mathbb{C}^{M\times1}, \label{signal}
\end{equation}
where $\mathbf{W}_k\in\mathbb{C}^{M\times N_{d}}$ denotes the transmit beamforming matrix corresponding to IDR $k$ with $N_d$ being the number of data streams for each user, and $\mathbf{s}_k \in \mathbb{C}^{N_{d} \times 1}$ is the data symbol vector of IDR $k$ satisfying $\mathbb{E}\{\mathbf{s}_k\mathbf{s}^\mathrm{H}_k\}=\mathbf{I}_{N_d}$ and $\mathbb{E}\{\mathbf{s}_j\mathbf{s}^{\mathrm{H}}_k\}=\mathbf{0}$, $\forall j\neq k$ with $\mathbf{I}$ and $\mathbf{0}$ denoting the identical and all-zero matrices, respectively. For simplicity, we assume $M\geq J$ and $N_d=\mathrm{min}\{M,J\}=J$.

Denote $h_{m,n,k,j}$ as the channel coefficient between the $n$-th PA on waveguide $m$ and the $j$-th antenna of IDR $k$, i.e.,
\begin{equation}
    h_{m,n,k,j}=\xi\frac{\mathrm{exp}\{-\jmath\kappa D_{m,k,j}(l_{m,n})\}}{D_{m,k,j}(l_{m,n})},
\end{equation}
where  $\xi=\frac{c}{4\pi f_c}$ and $\kappa=\frac{2\pi f_c}{c}$ with $c$ and $f_c$ representing the speed of light and carrier frequency, respectively. $D_{m,k,j}(l)$ is a function computing the distance between the position $l$ on the $m$-th waveguide and the $j$-th antenna of IDR $k$. 
Let the $m$-th element of the signal vector $\mathbf{s}$, i.e., $s_m$ denote the signal fed to waveguide $m$. The radiated signal from the $n$-th PA on waveguide $m$ is the attenuated and phase-shifted version of $s_m$ with the attenuation factor as $p_{m,n}$ and the phase-shift as $\kappa \imath_{\mathrm{ref}}l_{m,n}$, where $\imath_{\mathrm{ref}}$ is the refractive index of the waveguides. Assuming that the propagation loss of the signal $s_m$ within waveguide $m$ can be neglected, in this case the attenuation factor $p_{m,n}=\sqrt{1/N}$, $\forall n$ \cite{PASS1}.

Defining $\mathbf{l}_m=[l_{m,1},\cdots,l_{m,N}] \in \mathbb{R}^{1\times N}$ as the position vector of the waveguide $m$,  the effective channel coefficient between the waveguide $m$ and the $j$-th antenna of IDR $k$, i.e., $h_{m,k,j}(\mathbf{l}_m)\in\mathbb{C}$ is expressed as
\begin{align}
    h_{m,k,j}(\mathbf{l}_m)
    &=\xi\sum^N_{n=1}\frac{\mathrm{exp}\{-\jmath\kappa(D_{m,k,j}(l_{m,n})+\imath_{\mathrm{ref}}l_{m,n})\}}{\sqrt{N}D_{m,k,j}(l_{m,n})}. \label{h}
\end{align} Then, the signal received at the $j$-th antenna of IDR $k$ can be expressed as
\begin{equation}
    y^{\mathrm{I}}_{k,j} = \sum^M_{m=1}h_{m,k,j}(\mathbf{l}_m)s_m+z_{k,j},
\end{equation}
where $z_{k,j}$ is the additive Gaussian white noise (AWGN) with variance $\sigma^2$, i.e., $z_{k,j}\sim\mathcal{CN}(0,\sigma^2)$.
Defining $\mathbf{L}=[\mathbf{l}^{\text{T}}_1,\cdots,\mathbf{l}^{\text{T}}_M]^{\text{T}}\in\mathbb{R}^{M\times N}$ and $\mathbf{h}_{k,j}(\mathbf{L})=[h_{1,k,j}(\mathbf{l}_1), \cdots, h_{M,k,j}(\mathbf{l}_M)]^{\mathrm{T}}\in\mathbb{C}^{M\times 1}$, the signal vector received at IDR $k$ is hence given by
\begin{equation}
\mathbf{y}^{\mathrm{I}}_k=\mathbf{H}^{\mathrm{T}}_k(\mathbf{L})\mathbf{s}+\mathbf{z}_k, \label{yi}
\end{equation}
where $\mathbf{y}^{\mathrm{I}}_k=[y^{\mathrm{I}}_{k,1}, \cdots, y^{\mathrm{I}}_{k,J}]^{\mathrm{T}} \in \mathbb{C}^{J\times1}$ is the received signal vector, and the downlink MIMO channel matrix is defined as 
\begin{equation}
    \mathbf{H}_k(\mathbf{L})=[\mathbf{h}_{k,1}(\mathbf{L}), \cdots, \mathbf{h}_{k,J}(\mathbf{L})] \in \mathbb{C}^{M\times J} \label{H},
\end{equation}
and $\mathbf{z}_k=[z_{k,1}, \cdots, z_{k,J}]^{\mathrm{T}} \in \mathbb{C}^{J\times1}$ is the received AWGN satisfying $\mathbb{E}\{\mathbf{z}_k\mathbf{z}^{\mathrm{H}}_k\}=\sigma^2\mathbf{I}_J$. Similarly, the signal vector received at each EHR $q$ can be given by
\begin{equation}    \mathbf{y}^{\mathrm{E}}_q=\mathbf{G}^{\mathrm{T}}_q(\mathbf{L})\mathbf{s}+\mathbf{z}_q,
\end{equation}
where $\mathbf{G}_q(\mathbf{L})$ can be obtained similarly to \eqref{h} and \eqref{H}.

Based on \eqref{signal}, we can rewrite the signal received at IDR $k$ in \eqref{yi} as
\begin{equation}
    \mathbf{y}^{\mathrm{I}}_k=\mathbf{H}^{\mathrm{T}}_k(\mathbf{L})\mathbf{W}_k\mathbf{s}_k + \sum^K_{k'=1,k'\neq k}\mathbf{H}^{\mathrm{T}}_k(\mathbf{L})\mathbf{W}_{k'}\mathbf{s}_{k'}+\mathbf{z}_k,
\end{equation}
and the achievable data rate of IDR $k$ is given by
\begin{equation}
    R_k(\mathbf{W},\mathbf{L})=\mathrm{log}_2|\mathbf{I}_J+\mathbf{H}^{\mathrm{T}}_k(\mathbf{L})\mathbf{W}_k\mathbf{W}^{\mathrm{H}}_k\mathbf{H}^*_k(\mathbf{L})\mathbf{T}^{-1}_k|, \label{rate}
\end{equation}
where $\mathbf{T}_k=\sum\nolimits^K_{k'=1,k'\neq k}\mathbf{H}^{\mathrm{T}}_k(\mathbf{L})\mathbf{W}_{k'}\mathbf{W}^{\mathrm{H}}_{k'}\mathbf{H}^*_k(\mathbf{L})+\sigma^2\mathbf{I}_J$ is the interference-plus-noise-covariance matrix, and $\mathbf{W}=\{\mathbf{W}_k, \forall k\}$ is the collection of all transmit beamforming matrices. The energy harvested at EHR $q$ is given by \cite{MIMO}
\begin{equation}
    \mathcal{E}_q(\mathbf{W}, \mathbf{L}) \triangleq \eta_q\mathrm{Tr}\left(\sum^{K}_{k=1}\mathbf{G}^{\mathrm{T}}_q(\mathbf{L})\mathbf{W}_k\mathbf{W}^{\mathrm{H}}_k\mathbf{G}^*_q(\mathbf{L})\right),
\end{equation}
where $0<\eta_q<1$ is the efficiency of the energy
transducer to store the harvested energy. 

We aim to maximize the sum-rate of all IDRs by jointly optimizing the PA positions, $\mathbf{L}$, as well as the transmit beamforming matrices $\mathbf{W}$ while satisfying the constraints in terms of the minimum energy harvested at each EHR $q$, the transmission power, and the physical placement of PAs. The corresponding optimization problem is formulated as
\begin{subequations}
\begin{align}
&(\mathrm{P} 1): \max_{\mathbf{W},\mathbf{L}}~~
\sum^K_{k=1}\mathrm{log}_2|\mathbf{I}_J+\mathbf{H}^{\mathrm{T}}_k(\mathbf{L})\mathbf{W}_k\mathbf{W}^{\mathrm{H}}_k\mathbf{H}^*_k(\mathbf{L})\mathbf{T}^{-1}_k| \notag \\
&\text { s.t. } \notag \\ &\eta_q\mathrm{Tr}\left(\sum^{K}_{k=1}\mathbf{G}^{\mathrm{T}}_q(\mathbf{L})\mathbf{W}_k\mathbf{W}^{\mathrm{H}}_k\mathbf{G}^*_q(\mathbf{L})\right) \geq E_{\mathrm{min}}, \forall q, \label{a}\\
&  \sum^K_{k=1}\mathrm{Tr}(\mathbf{W}_k\mathbf{W}^{\mathrm{H}}_k) \leq P_{\mathrm{max}}, \label{b}\\
&  0 \leq l_{m, n} \leq L_m, \forall m, n \label{c}\\
&  l_{m, n}-l_{m, n-1} \geq L_0, \forall m, \forall n=2, \cdots, N, \label{d}
\end{align}
\end{subequations}
where \eqref{a} denotes that the harvested energy at each EHR $q$ should be higher than the minimum energy threshold $E_{\mathrm{min}}$, $P_{\mathrm{max}}$ denotes the maximum transmit power budget, \eqref{c} restricts the positions of PAs, and \eqref{d} guarantees the minimum spacing between adjacent PAs to prevent mutual coupling. 
\section{Proposed Algorithm}
Problem (P1) is a highly non-convex and NP-hard problem due to the non-convex objective function and constraints. To efficiently solve (P1), we introduce an AO-based algorithm. In the inner loop, we aim to iteratively optimize the transmit beamforming matrix $\mathbf{W}$ using a WMMSE-based method in a block coordinate descent (BCD) manner. In the outer loop, we aim to maximize the objective function using an iterative grid search method.
\subsubsection{Inner Loop Update}
Assuming that the PA position matrix $\mathbf{L}$ is fixed, we omit $\mathbf{L}$ for clarity in this section. To handle the non-convex objective in (P1), we apply the relationships between the WMMSE and the data rate expression in \eqref{rate} \cite{wmmse}. 

We consider a set of linear receive filters $\mathbf{U} \triangleq \{\mathbf{U}_k,\forall k\}$ for all IDRs so that the estimated signal vector of $\mathbf{s}_k$, i.e., $\hat{\mathbf{s}}_k$ is given by
\begin{equation}
    \hat{\mathbf{s}}_k=\mathbf{U}^{\mathrm{H}}_k\mathbf{y}^{\mathrm{I}}_k \in \mathbb{C}^{J \times 1}.
\end{equation}
Then, the mean-squared-error matrix $\mathbf{V}_k \in \mathbb{C}^{J\times J}$ between $\hat{\mathbf{s}}_k$ and $\mathbf{s}_k$ can be written as
\begin{align}
    \mathbf{V}_k &\triangleq \mathbb{E}\{(\hat{\mathbf{s}}_k-\mathbf{s}_k)(\hat{\mathbf{s}}_k-\mathbf{s}_k)^{\mathrm{H}}\} \notag \\
    &=(\mathbf{U}^{\mathrm{H}}_k \mathbf{H}^{\mathrm{T}}_k\mathbf{W}_k-\mathbf{I}_J)(\mathbf{U}^{\mathrm{H}}_k \mathbf{H}^{\mathrm{T}}_k\mathbf{W}_k-\mathbf{I}_J)^{\mathrm{H}} \notag \\
    &~ + \sum^K_{k'=1, k'\neq k} \mathbf{U}^{\mathrm{H}}_k \mathbf{H}^{\mathrm{T}}_k \mathbf{W}_{k'}\mathbf{W}^{\mathrm{H}}_{k'}\mathbf{H}^{\mathrm{*}}_k\mathbf{U}_k + \sigma^2 \mathbf{U}^{\mathrm{H}}_k\mathbf{U}_k. \label{mse}
\end{align}
By introducing a set of auxiliary variables $\boldsymbol{\Lambda}\triangleq\{\boldsymbol{\Lambda}_k, \forall k\}$, we define the surrogate function in terms of the data rates of IDR $k$ as follows
\begin{equation}
f_k(\mathbf{W},\mathbf{U}_k,\boldsymbol{\Lambda}_k)\triangleq \mathrm{log}_2|\boldsymbol{\Lambda}_k|-\mathrm{Tr}(\boldsymbol{\Lambda}_k\mathbf{V}_k)+J, ~\forall k. \label{fk0}
\end{equation}
From Lemma 1 in \cite{pan}, it can be concluded that $f_k(\mathbf{W},\mathbf{U}_k,\boldsymbol{\Lambda}_k)$ is a concave function for each set of $\mathbf{W}$, $\mathbf{U}_k$ and $\boldsymbol{\Lambda}_k$, and is upper bounded by the data rate $R_k(\mathbf{W})$. The optimal $\mathbf{U}_k$, $\boldsymbol{\Lambda}_k$, and $\mathbf{V}_k$ for $f_k(\mathbf{W},\mathbf{U}_k,\boldsymbol{\Lambda}_k)$ to achieve the data rate $R_k(\mathbf{W})$ are given by
\begin{align}
    \mathbf{U}^{\mathrm{opt}}_k &= \left(\sum^K_{k'=1}\mathbf{H}^{\mathrm{T}}_k\mathbf{W}_{k'}\mathbf{W}^{\mathrm{H}}_{k'}\mathbf{H}^{\mathrm{*}}_k+\sigma^2\mathbf{I}_J\right)^{-1}\mathbf{H}^{\mathrm{T}}_k\mathbf{W}_k, \label{uo}\\
    \boldsymbol{\Lambda}^{\mathrm{opt}}_k &= (\mathbf{V}^{\mathrm{opt}}_k)^{-1}, \label{lamo}\\
    \mathbf{V}^{\mathrm{opt}}_k &= \mathbf{I}_J - \mathbf{W}^{\mathrm{H}}_k\mathbf{H}^{\mathrm{*}}_k\times \notag \\
    &\quad \left(\sum^K_{k'=1}\mathbf{H}^{\mathrm{T}}_k\mathbf{W}_{k'}\mathbf{W}^{\mathrm{H}}_{k'}\mathbf{H}^{\mathrm{*}}_k+\sigma^2\mathbf{I}_J\right)^{-1}\mathbf{H}^{\mathrm{T}}_k\mathbf{W}_k. \label{vo}
\end{align}

By substituting the expression of $\mathbf{V}_k$ in \eqref{mse} into the surrogate function in \eqref{fk0}, we can rewrite it as in \eqref{fk}. The objective function in (P1) is converted to
\setcounter{equation}{20}
\begin{align}
    \mathcal{F}(\mathbf{W})=\sum^K_{k=1}R_k(\mathbf{W}) \geq \sum^K_{k=1}f_k(\mathbf{W},\mathbf{U}_k,\boldsymbol{\Lambda}_k). \label{newa}
\end{align}
\begin{figure*}[!t]
\vspace{-1em}
\normalsize
\setcounter{equation}{19}
  \begin{align}
      f_k(\mathbf{W},\mathbf{U}_k,\boldsymbol{\Lambda}_k) = \mathrm{log}_2|\boldsymbol{\Lambda}_k|&-\mathrm{Tr}\left(\boldsymbol{\Lambda}_k\sum^K_{k'=1}\mathbf{U}^{\mathrm{H}}_k\mathbf{H}^{\mathrm{T}}_k\mathbf{W}_{k'}\mathbf{W}^{\mathrm{H}}_{k'}\mathbf{H}^{\mathrm{*}}_k\mathbf{U}_k\right)+\mathrm{Tr}\bigg(\boldsymbol{\Lambda}_k\mathbf{U}^{\mathrm{H}}_k\mathbf{H}^{\mathrm{T}}_k\mathbf{W}_k\bigg) \notag \\
      &~+ \mathrm{Tr}\bigg(\boldsymbol{\Lambda}_k\mathbf{W}^{\mathrm{H}}_k\mathbf{H}^*_k\mathbf{U}_k\bigg)-\mathrm{Tr}(\boldsymbol{\Lambda}_k)-\mathrm{Tr}(\sigma^2\boldsymbol{\Lambda}_k\mathbf{U}^{\mathrm{H}}_k\mathbf{U}_k)+J \label{fk}
  \end{align}
  \setcounter{equation}{21}
\begin{align}
\mathcal{E}_q(\mathbf{W})
\geq \tilde{\mathcal{E}}_q(\mathbf{W}, \mathbf{W}^{(t)}) \triangleq\eta_q\bigg[-\mathrm{Tr}\left(\sum^{K}_{k=1}\mathbf{W}^{(t),\mathrm{H}}_k\mathbf{G}^*_q\mathbf{G}^{\mathrm{T}}_q\mathbf{W}^{(t)}_k\right)
&+2\mathrm{Re}\bigg\{ \mathrm{Tr}\left(\sum^{K}_{k=1}\mathbf{W}^{(t),\mathrm{H}}_k\mathbf{G}^*_q\mathbf{G}^{\mathrm{T}}_q\mathbf{W}_k\right)\bigg\} \bigg], ~\forall q. \label{fw}
\end{align}
  \noindent\rule{\textwidth}{0.4pt}
\end{figure*}
After the transformation of the objective function, we utilize the successive convex approximation (SCA) technique to transform the constraints in \eqref{a} into \eqref{fw},
where $\mathbf{W}^{(t)}\triangleq\{\mathbf{W}^{(t)}_k, \forall k\}$ is the set of beamforming matrices optimized at the $t$-th iteration in the inner loop. Hence, problem (P1) is transformed into 
\begin{subequations}
    \begin{align}
        (\mathrm{P} 2): &\max_{\mathbf{W},\mathbf{U}, \boldsymbol{\Lambda}}~~
\sum^K_{k=1} f_k(\mathbf{W},\mathbf{U}_k,\boldsymbol{\Lambda}_k) \notag \\
\text { s.t. } \notag \\ &\tilde{\mathcal{E}}_q(\mathbf{W}, \mathbf{W}^{(t)}) \geq E_{\mathrm{min}}, ~\forall q, \label{aa}\\
&  \sum^K_{k=1}\mathrm{Tr}(\mathbf{W}_k\mathbf{W}^{\mathrm{H}}_k) \leq P_{\mathrm{max}}. \label{bb}
    \end{align}
\end{subequations}
It is clear that the objective function of problem (P2) is concave, and the constraints \eqref{aa} and \eqref{bb} are all convex with respect to $\mathbf{W}$. Therefore, problem (P2) is a quadratically constrained quadratic programming (QCQP) problem and can be easily solved via some standard solvers like CVX.
\subsubsection{Outer Loop Update}
At the $(I+1)$-th iteration of the outer loop, we aim to optimize the PA position matrix $\mathbf{L}$ with the fixed optimized beamforming matrix $\mathbf{W}$, the receive filter $\mathbf{U}$, and the auxiliary variable $\boldsymbol{\Lambda}$. We use an iterative Gauss-Seidel-based method, in which each PA position $l_{m,n}, \forall m,n$ is individually updated while treating other positions as constants.

Consider the $n$-th element on waveguide $m$ and fix $l_{m',n'}$, $\forall (m',n')\neq(m,n)$. We define the function $\Pi_{m,k,j}(l)$ as
\begin{equation}
    \Pi_{m,k,j}(l)=\xi\frac{\mathrm{exp}\{-\jmath\kappa(D_{m,k,j}(l)+\imath_{\mathrm{ref}}l) \}}{\sqrt{N}D_{m,k,j}(l)},
\end{equation}
where $l$ is the PA position.
By defining the $(m,n)$-th entry of $\{\mathbf{W}_{k'}\mathbf{W}^{\mathrm{H}}_{k'}\}, \forall k'$ as $\tilde{w}_{k,m,n}$, the $(i,j)$-th entry of $\{\mathbf{U}_k\boldsymbol{\Lambda}_k\mathbf{U}^{\mathrm{H}}_k\},\forall k$ as $\tilde{\Lambda}_{k,i,j}$, and that of $\{\mathbf{W}_k\boldsymbol{\Lambda}_k\mathbf{U}^{\mathrm{H}}_k\},\forall k$ as $\ddot{\Lambda}_{k,i,j}$, we can re-express the second term in \eqref{fk} as follows by simple lines of derivation 
\begin{align}
\hat{\Pi}_{k}(l_{m,n})&\triangleq\sum^J_{i=1}\sum^K_{k'=1}\sum^J_{j=1}\tilde{\Pi}_{k',k,i,j}(l_{m,n})\tilde{\Lambda}_{k,j,i},
\end{align}\
where
\begin{align}
    &\tilde{\Pi}_{k',k,i,j}(l_{m,n}) 
    \triangleq\sum^M_{m'=1,m'\neq m}\Pi_{m,k,i}(l_{m,n})\tilde{w}_{k',m,m'}h^*_{m',k,j}(\mathbf{l}_{m'}) \notag \\
    &+\sum^M_{m'=1,m'\neq m}\Pi^*_{m,k,j}(l_{m,n})\tilde{w}_{k',m',m}h_{m',k,i}(\mathbf{l}_{m'}) \notag \\
    &+\sum^N_{n'=1,n'\neq n}\Pi_{m,k,i}(l_{m,n'})\Pi^*_{m,k,j}(l_{m,n})\tilde{w}_{k',m,m} \notag \\
    &+\sum^N_{n'=1,n'\neq n}\Pi^*_{m,k,j}(l_{m,n'})\Pi_{m,k,i}(l_{m,n})\tilde{w}_{k',m,m} \notag \\
    &+\Pi_{m,k,i}(l_{m,n})\Pi^*_{m,k,j}(l_{m,n})\tilde{w}_{k',m,m}+\Delta_{k',k,i,j}, \label{pi}
\end{align}
where $\Delta_{k',k,i,j} \in \mathbb{C}$ is a constant term, regardless of $l_{m,n}$. 
Then, we can re-express the third term in \eqref{fk} as
\begin{align}
    \ddot{\Pi}_{k}(l_{m,n}) &\triangleq \sum^J_{j=1}\Pi_{m,k,j}(l_{m,n})\ddot{\Lambda}_{k,j,j},
\end{align}
and rewrite \eqref{a} as 
\begin{equation}
\sum^J_{i=1}\sum^K_{k=1}\eta_q\tilde{\Pi}_{k,q,i,i}(l_{m,n}) \geq E_{\mathrm{min}},~ \forall q, \label{ddd} 
\end{equation}
where $\tilde{\Pi}_{k,q,i,i}(l_{m,n})$ can be derived similarly as in \eqref{pi}.

Hence, the scalar problem of optimizing $l_{m,n}$ is equivalently written as
\begin{subequations}  
    \begin{align}
(\mathrm{P} 3):
\max_{l_{m,n}} ~&\sum^K_{k=1} 2\mathrm{Re}\left\{\ddot{\Pi}_{k}(l_{m,n})\right\}-\hat{\Pi}_{k}(l_{m,n}) \notag \\
\text { s.t. }
 &~l_{m,n'}-l_{m,n'} \geq L_0, ~\forall n'=n,n+1, \label{d2} \\
  &~\eqref{c} ~\text{and} ~\eqref{ddd}. \notag
\end{align}
\end{subequations}
Problem (P3) can be effectively solved via a one-dimensional grid search with a search interval within $[0, L_m]$ consisting of $L$ search points, which is given by $\mathbb{G}_m=\{0, \frac{L_m}{L-1},\frac{2L_m}{L-1},\cdots,L_m \}.$

\subsubsection{Algorithm Summary}
The overall algorithm for solving problem (P1) is summarized in Algorithm \ref{alg}. The updates of $\mathbf{W}$ and $\mathbf{L}$ result in a non-decrease in the objective function. Moreover, the feasible regions of $\mathbf{W}$ and $\mathbf{L}$ are bounded. Hence, Algorithm \ref{alg} is guaranteed to converge. The complexity of Algorithm \ref{alg} is dominated by solving the QCQP problem (P2), which is $\mathcal{O}(M^3J^3K^3)$. 
\begin{algorithm}[t]
\footnotesize
    \caption{Algorithm for Solving Problem (P1)}
    \label{alg}
    \textbf{Initialize:} Outer iteration number $I=0$, inner iteration number $t=0$, $\mathbf{W}^{(t)}$, $\mathbf{L}^{(I)}$ and $L$. \\
   \textbf{Outer loop:}\\
    \Repeat{convergence}{
    {\textbf{Inner loop:}\\
    \Repeat{convergence}{Update $\mathbf{U}^{(t+1)}$ and $\boldsymbol{\Lambda}^{(t+1)}$ based on \eqref{uo} and \eqref{lamo}.\\
    Update $\mathbf{W}^{(t+1)}$ by solving problem (P2).\\
    $t=t+1$.\\}
    }
    Update $\mathbf{L}^{(I+1)}$ by iteratively solving problem (P3).\\

    $I=I+1$.
    }
    \Return{$\mathbf{W}^{\mathrm{opt}}$} and $\mathbf{L}^{\mathrm{opt}}$.
\end{algorithm}

\section{Numerical Results}
This section evaluates the performance of the proposed MIMO PASS multi-user SWIPT system. Specifically, we consider a rectangular region with $L_x = 30$ m and $L_y=6$ m.  Within this region, $K=2$ IDRs and $Q=2$ EHRs are randomly and uniformly distributed. Other system setting parameters are shown as follows: $f_c=28$ GHz, $J=3$, $d_s=\lambda/2$, $M=4$, $a=5$ m, $d=L_y/(M-1)$, $N=3$, $L_0=\lambda/2$, $\imath_{\mathrm{ref}}=1.44$, $\sigma^2=-50$ dBm,  $P_{\mathrm{max}}=43$ dBm,  $E_{\mathrm{min}}=0.5\times10^{-7}$ W, $L=2001$, and $\eta_q=50\%$. We consider the following benchmark schemes. 
\textit{1):} \textbf{PASS with ZF design}: In this case, a zero-forcing (ZF)-based beamforming method is utilized instead of the proposed design. Then, we solve problem (P2) to optimize the PA position $\mathbf{L}$; \textit{2):} \textbf{Fixed PA position design}: In this case, PAs are uniformly distributed on each waveguide with fixed positions $l_{m,n}=\frac{L_x}{N+1}\times n$. Then we solve problem (P1) to optimize the pinching beamforming $\mathbf{W}$; \textit{3):} \textbf{Conventional MIMO design}: In this case, we consider a MIMO system, where the AP is equipped with a half-wavelength-spaced ULA centered within the rectangular region at $[L_x/2, L_y/2, a]$ with $M$ antennas, each connected to a dedicated RF chain. 
\begin{figure}[h]
    \centering
    \includegraphics[width=0.6\linewidth]{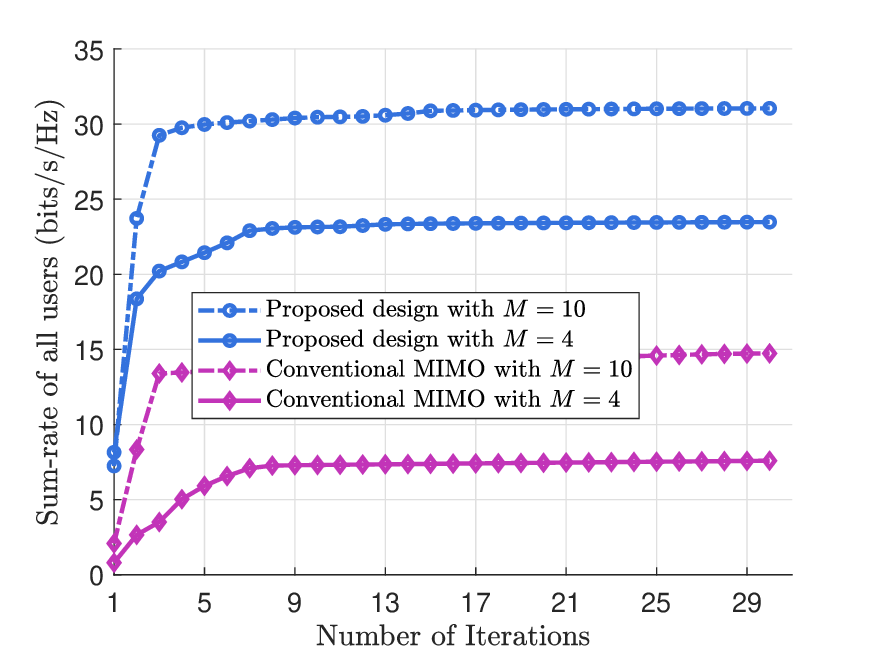}
    \caption{Convergence of the proposed design.}
    \label{conv}
    \vspace{-10pt}
\end{figure}

First, Fig. \ref{conv} illustrates the convergence behavior of the proposed design with different numbers of waveguides compared with conventional MIMO systems. As observed, the sum-rate of the proposed design increases rapidly with the increase of iterations and converges within 20 iterations for all considered values of $M$. Moreover, the proposed design significantly outperforms the conventional MIMO system. Despite the increase in the number of MIMO antennas, the performance of the conventional MIMO system remains inferior to that of the proposed design with $M=4$ waveguides. This verifies that the PASS are able to explore more spatial DoFs to enhance the performance of SWIPT than conventional MIMO systems.

\begin{figure}[h]
    \centering
    \includegraphics[width=0.6\linewidth]{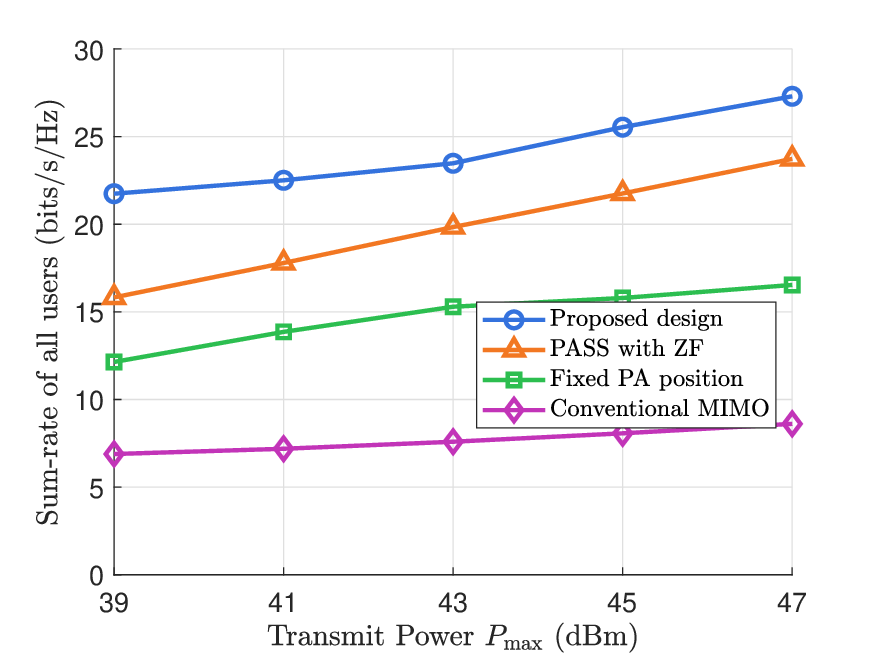}
    \caption{Sum-rate as a function of transmit power $P_{\mathrm{max}}$.}
    \label{power}
\end{figure}

Next, Fig. \ref{power} demonstrates the sum-rate as a function of transmit power $P_{\mathrm{max}}$. It is observed that the sum-rate of all schemes increases with the transmit power because more power is allocated for transmitting information to IDRs, and the EH requirements are easier to satisfy. Compared with the Fixed PA position design and the Conventional MIMO design, the proposed scheme achieves a huge performance gap. This demonstrates the effectiveness of PASS in adjusting PA positions to acquire more spatial DoFs to compensate for the large-scale path loss and enhance the SWIPT performance. Moreover, as the transmit power increases, the ZF-based design outperforms the Fixed PA position design and gradually approaches the proposed design. This is because more transmit power leads to less severe path loss, which creates a relatively high-SNR scenario where ZF performs well.

\begin{figure}[h]
    \centering
    \includegraphics[width=0.6\linewidth]{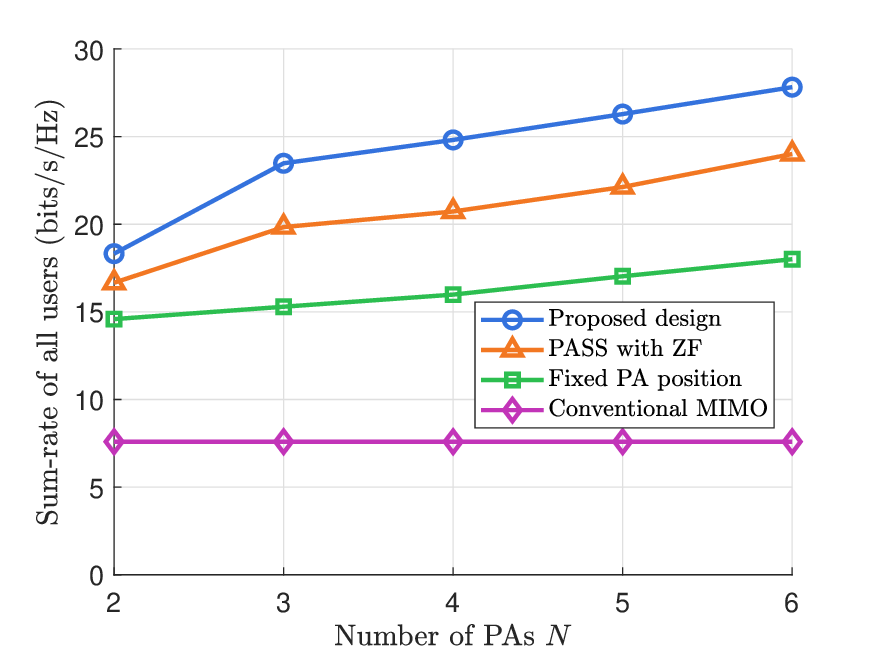}
    \caption{Sum-rate as a function of the number of PAs per waveguide $N$.}
    \label{N}
\end{figure}

Finally, Fig. \ref{N} shows the sum-rate as a function of the number of PAs per waveguide $N$. It is shown that the sum-rate of the proposed design increases with the number of PAs due to more spatial DoFs and beamforming gains, which mitigates the inter-IDR interference and improves the power transfer efficiency. Moreover, compared with the conventional MIMO design, the Fixed PA position design still demonstrates huge performance superiority with the increase of the number of PAs, since more PAs are spread over the waveguides and closer to IDRs and EHRs, resulting in strengthened LoS links.
\section{Conclusions}
This letter has introduced a novel MIMO PASS-aided multi-user SWIPT system, where the pinching beamforming and the PA positions were jointly optimized to enhance the performance of information transmission and WPT. By invoking the WMMSE-based method and the Gauss-Seidel approach, we developed an AO framework to efficiently solve the highly non-convex optimization problem. Numerical results verified the superior performance of our proposed design compared with various benchmark schemes.

\end{document}